\documentclass[12pt,a4paper]{article}

\DeclareSymbolFont{AMSb}{U}{msb}{m}{n}
\DeclareSymbolFontAlphabet{\Bbb}{AMSb}

\newcommand{\R}{\Bbb{R}}

\newtheorem{theorem}{Theorem}[section]

\newtheorem{lemma}[theorem]{Lemma}
\newtheorem{proposition}[theorem]{Proposition}
\newtheorem{definition}[theorem]{Definition}
\newtheorem{example}[theorem]{Example}

\newcommand{\shp}{\mathcal{C}}
\newcommand{\shq}{\mathcal{D}}
\newcommand{\unp}{\mathcal{F}}
\newcommand{\unq}{\mathcal{G}}
\newcommand{\idty}{\mathbf{1}}
\newcommand{\Tr}{\mathrm{Tr}\,}
\newcommand{\alg}{\mathcal{A}}
\newcommand{\algg}{\mathcal{B}}
\newcommand{\matr}[1]{\mathcal{M}_{#1}}
\newcommand{\ctu}{\mathcal{C}}
\newcommand{\oa}{\overline{\alpha}}
\newcommand{\ok}{\overline{k}}
\newcommand{\ol}{\overline{l}}
\newcommand{\h}[1]{h^{\mathrm{#1}}}
\newcommand{\HH}[1]{H^{\mathrm{#1}}}
\newcommand{\ro}[1]{\rho^{\mathrm{#1}}}
\newcommand{\Hrel}{I}
\newcommand{\qed}{\hfill qed}

\begin{document}
\bigskip
\begin{center}
\textbf{\LARGE Quantum Dynamical Entropies \\*[2mm]
for Classical Stochastic Systems}\\*[15mm]
{\large Mark Fannes and Bart Haegeman
\footnote{Aspirant FWO-Vlaanderen}}
\\*[5mm]
Instituut voor Theoretische Fysica \\
K.U. Leuven, B-3001 Leuven, Belgium 
\end{center}
\bigskip
\bigskip
\begin{abstract}
\noindent 
We compare two proposals for the dynamical entropy of quantum
deterministic systems (CNT and AFL) by studying their extensions
to classical stochastic systems. We show that the natural measurement
procedure leads to a simple explicit expression for the stochastic
dynamical entropy with a clear information-theoretical interpretation.
Finally, we compare our construction with other recent proposals.
\end{abstract}

\section{Motivation}

Dynamical entropy is a standard tool for the study of classical
deterministic systems, see, e.g., \cite{W}. It measures the
marginal amount of uncertainty generated by the dynamics, or, 
equivalently, the marginal amount of information obtained about
the initial condition. Different approaches have been followed to
generalize the idea of dynamical entropy to quantum systems,
\cite{AF1,AF2,AOW,CNT,H}. Thereby, one encounters, apart from the
non-commutativity, still another problem~: Not only the dynamics can
generate uncertainty, but also the quantum measurements can do so.
A good calculation scheme should separate these two contributions.

The latter problem is also present when extending the notion of dynamical
entropy to classical stochastic systems. In this case, the different
sources of entropy production to separate are the system dynamics and
the stochasticity due to the coupling to the unobserved environment. 
Therefore, from the dynamical entropy point of view, classical stochastic
dynamics can be considered as an intermediate case between classical 
deterministic and quantum dynamics, \cite{M}.

We will make this link even more explicit by taking two established
quantum constructions (CNT \cite{CNT} and AFL \cite{AF1, AF2}) as
a starting point and extending them to stochastic systems.
As a consequence, next to system and environment, so-called unsharp
measurements appear as a third source of dynamical entropy.
Interestingly, the different quantum constructions lead to 
clear-cut differences which we can interpret in terms of these three 
sources. E.g., for the extreme situation of a Bernoulli process, where 
the stationary completely random state is already reached after a
single time step, successive observations of the system do not reveal
any information at all on the initial state of our system. The degree
of stochasticity of the dynamics can be measured however and,
moreover, very unsharp observations of such a process will overestimate 
this randomness.

Quantum dynamical entropy has recently received new interest in
connection with quantum information theory. Both dynamical entropies
we will discuss in this paper have been reformulated in this
framework, CNT in \cite{B} and AFL in \cite{A}. The stochasticity
we introduce can both model badly isolated information sources
or noisy communication channels. Finally, this work can also be
considered as a first step in the construction of a dynamical entropy
for quantum stochastic systems, as was recently done in \cite{KOW}.

\section{Preliminaries}

Deterministic classic dynamics are given by a transformation $T$ of the
phase space $X$. For stochastic systems one should use stochastic
transformations of phase space. It is more convenient to work on the level of
the observables, i.e., the functions on the phase space. Such a description
allows also to connect with the quantum world by allowing the algebra of
observables to become non-commutative.  The dynamics is now given by a
transformation $\Theta$ of a function space on the phase space $X$. The
different concepts needed to introduce dynamical entropy, like partitions,
their evolutions and refinements, must be transported from the level of the
points of phase space to the level of observables.

\subsection{Some notation}

Let $\mu$ be a probability measure on the set $X$. Consider a transformation
$\Theta$ of the algebra of observables, $\Theta: L^\infty(X,\mu)\to
L^\infty(X,\mu)$, which is
\begin{itemize} 
\item 
 positive, $f\geq 0\Rightarrow \Theta(f)\geq 0$, for all $f\in
 L^\infty(X,\mu)$,
\item 
 unital, $\Theta(\idty)=\idty$ and
\item 
 measure-preserving, $\mu(\Theta(f))=\mu(f)$, for all $f\in
 L^\infty(X,\mu)$.
\end{itemize}
The triple $(X,\mu,\Theta)$ defines a {\em stochastic dynamical system} in
discrete time.

\begin{example}[Markov process] \label{exam:markov}
Let $X$ be a finite set and
$\mu=\{\mu_x\,|\,x\in X\}$ a probability measure on $X$.
Let $P$ be a transition matrix satisfying
\[
 P_{xy}\geq 0, \quad \sum_y P_{xy}=1
 \quad\mathrm{and}\quad \sum_x\mu_x P_{xy}=\mu_y.
\]
The time evolution $\Theta$, given by
\[
 \Theta(f)(x) := \sum_y P_{xy}f(y),
\]
defines a stochastic dynamical system.
This finite-dimensional example can be generalized considerably.
Let $(X,\mathcal{S},\mu)$ be a $\sigma$-finite probability space.
Let $P$ be a measurable function on the product space
$X\times X$ satisfying
\[
 P(x,y)\geq 0, \quad \int d\mu(y)P(x,y)=1,\ x\in X
\]
and
\[
 \int d\mu(x)\int_S d\mu(y)P(x,y)=\mu(S),\ S\in\mathcal{S}.
\]
The time evolution $\Theta$ is given by
\[
 \Theta(f)(x) := \int d\mu(y)P(x,y)f(y).
\]
\end{example}

\begin{example}[Deterministic systems] \label{exam:determ}
Also deterministic dynamical systems are included in this formalism. They
are given by a probability space $(X,\mathcal{S},\mu)$ and a transformation
$T:X\to X$ which is measure-preserving, $\mu(T^{-1}(S))=\mu(S),\
S\in\mathcal{S}$. Take then $\Theta(f) := f\circ T$.
\end{example}

The positivity of $\Theta$ can be rephrased as 
$|\Theta(f)| \leq \Theta(|f|)$ (triangle inequality) or as
$|\Theta(f)|^2 \leq \Theta(|f|^2)$ (Schwarz inequality).
Deterministic systems are then distinguished by the additional property
$|\Theta(f)| = \Theta(|f|)$ or by $|\Theta(f)|^2 = \Theta(|f|^2)$.
This means exactly that $\Theta$ is an endomorphism of $L^\infty(X,\mu)$.

We are ready now to construct a partition on the level of the observables. A
set of measurable functions $\unp=\{f_k\,|\,k\in K\}$ with $K$ a finite index
set, is called a {\em partition of unity} whenever $f_k\geq 0$ and 
$\sum_kf_k=\idty$. Such a set of functions can be interpreted as a response
function for an unsharp measurement. The number $f_k(x)$ equals the probability
for the measurement outcome $k\in K$ given the system is located in $x\in X$.

\begin{example}[Sharp measurements] \label{exam:sharp}
An important class of partitions of unity are those corresponding to sharp
measurements. Let $\shp=\{C_k\,|\,k\in K\}$ be a measurable partition of
$X$, i.e., $C_k\subset X$ is measurable, $C_k\cap C_l=C_k\delta_{kl}$ and
$\bigcup_kC_k=X$. The set
\[
 \chi_\shp := \{\chi_{C_k}\,|\,k\in K\},
\]
where $\chi_C$ denotes the characteristic function of the set $C\subset X$,
is a partition of unity.
\end{example}

Because deterministic dynamics act on the level of points of phase space,
sharp measurements suffice in this case. Stochastic dynamics, on the
contrary, smooth out sharp measurements. It is then natural to consider
unsharp measurements. This implies, however, that a measurement as such,
i.e., independent of the dynamics, can contribute to the dynamical entropy.
Compared to classical deterministic systems, this is a new phenomenon one
should take care off when constructing a dynamical entropy.

\subsection{Refined partitions of unity}

For two partitions of unity,
$\unp=\{f_k\,|\,k\in K\}$ and $\unq=\{g_l\,|\,l\in L\}$,
let $\unp\lor\unq$ denote the partition of unity
$\{f_kg_l\,|\,k\in K,\ l\in L\}$.
Also, define the time evolution of a partition of unity $\unp$,
$\Theta(\unp)=\{\Theta(f_k)\,|\,k\in K\}$,
which is again a partition of unity.

The repeated application of the time evolution $\Theta$
can be described as a refinement of an initial partition
of unity. For deterministic dynamics, for example,
the refined partition of unity after $n$ time steps is given by
\begin{equation}
\unp\lor\Theta(\unp)\lor\ldots\lor\Theta^{n-1}(\unp).
\label{eq:refineMak}
\end{equation}
This can be used as a definition for the evolution of
a partition of unity under stochastic dynamics, \cite{M}.
In this paper, we will use another definition,
\begin{equation}
\unp^{(n)}[\Theta]:=\unp\lor\Theta(\unp\lor\ldots\lor\Theta(\unp)),
\label{eq:refineAF}
\end{equation}
where the initial partition of unity $\unp$ appears $n$ times.
As $\Theta$ is generally only positive and not necessarily an endomorphism,
the definitions (\ref{eq:refineMak}) and (\ref{eq:refineAF})
coincide for deterministic systems, but differ for stochastic
systems. This difference can be illustrated for a Markov process,
Example~\ref{exam:markov}.

\begin{example}[Markov process]
Let $X$ be the finite state space and $\unp=\{f_k\}$ a partition
of unity. The element $(k_0,k_1,\ldots,k_{N-1})$ of the refined
partition of unity (\ref{eq:refineAF}) is given by
\[
x \mapsto \sum_{(x_0,x_1,\ldots,x_{N-1})}
f_{k_0}(x)P_{xx_1}f_{k_1}(x_1)P_{x_1x_2} \ldots
P_{x_{N-2}x_{N-1}}f_{k_{N-1}}(x_{N-1}),
\]
and equals the probability for the measurement outcome
$(k_0,k_1,\ldots,k_{N-1})$ given the initial state $x\in X$.
A similar interpretation is missing for the refinement
(\ref{eq:refineMak}),
\[
x \mapsto \sum_{(x_0,x_1,\ldots,x_{N-1})}
f_{k_0}(x)\,P_{xx_1}f_{k_1}(x_1)\,P^{(2)}_{xx_2}f_{k_2}(x_2)
\ldots P^{(N-1)}_{xx_{N-1}}f_{k_{N-1}}(x_{N-1}),
\]
where $P^{(n)}$ is the $n$-th matrix power of $P$.
\end{example}

In quantum systems, partitions are replaced by so-called operational
partitions, i.e., sets of observables $\mathcal{X}=\{x_k \mid k\in K\}$
such that $\sum_k x_k^*x_k = \idty$. For two such partitions,
$\mathcal{X}=\{x_k \mid k\in K\}$ and $\mathcal{Y}=\{y_l \mid l\in L\}$,
we can again define
$\mathcal{X}\lor\mathcal{Y}=\{x_ky_l\,|\,k\in K,l\in L\}$ and
$\Theta(\mathcal{X})=\{\Theta(x_k)\,|\,k\in K\}$.
For endomorphisms $\Theta$ the refinement of $\mathcal{X}$ can then
be defined as in Eq.~(\ref{eq:refineMak}) or, equivalently,
Eq.~(\ref{eq:refineAF}). However, for non-endomorphic maps $\Theta$
this approach does not work because $\Theta(\mathcal{X})$ will not
be an operational partition anymore. Instead, at time $n$
the $\#(K)$ operators $x_k^*x_k$ should be replaced by
\[
x_{k_0}^* \Theta(x_{k_1}^* \ldots
\Theta(x_{k_{n-1}}^*x_{k_{n-1}})
\ldots x_{k_1}) x_{k_0},
\]
which are $\#(K)^n$ positive operators summing up to $\idty$.
This is a generalization of (\ref{eq:refineAF}) rather than of
(\ref{eq:refineMak}).

\subsection{Entropy and all that}

In this subsection we collect for finite systems the definitions and properties
of entropy and of relative entropy which will be needed  later~\cite{We,OP}.
These properties naturally extend to infinite systems.

With $\eta:[0,1]\to\R$ the entropy function,
\[
\begin{array}{llll}
\eta(x) & := & -x\log x, & x\in(0,1] \\
        & := & 0,       & x=0,
\end{array}
\]
we have the following definitions.

\begin{definition}
The entropy (or Shannon entropy) of a probability measure
$\mu=\{\mu_i\,|\,i\in I\}$ with finite index set $I$, is given by,
\[
 S(\mu) := \sum_{i\in I}\eta(\mu_i).
\]
The quantum entropy (or von Neumann entropy) of a density matrix $\rho$
on a finite-dimensional Hilbert space is given by,
\[
 S_q(\rho) := \Tr\eta(\rho).
\]
\end{definition}

If we denote by $\mathrm{diag}(\rho)$ the probability measure
obtained by restricting the density matrix $\rho$ to its diagonal
in a given basis, then $S_q(\rho)\leq S(\mathrm{diag}(\rho))$.
Because the entropy function $\eta$ is concave, the entropies
$S$ and $S_q$ are also concave. Moreover, one has,

\begin{proposition}[Concavity of entropy] \label{prop:concave}
Let $\mu$, $\mu_\alpha$, $\alpha\in A$ and $\#(A)<\infty$,
be probability measures on a finite set $I$ and
let $\lambda=\{\lambda_\alpha\,|\,\alpha\in A\}$ be
a probability measure such that
\[
\mu=\sum_{\alpha\in A}\lambda_\alpha\mu_\alpha,
\]
then
\[
\sum_\alpha\lambda_\alpha S(\mu_\alpha) \leq S(\mu) \leq
\sum_\alpha\lambda_\alpha S(\mu_\alpha) + S(\lambda).
\]
Similarly, for a convex combination of density matrices
$\rho_\alpha$ on a finite-dimensional Hilbert space,
\[
\rho=\sum_\alpha\lambda_\alpha\rho_\alpha,
\]
it holds that
\begin{equation} \label{eq:concave}
\sum_\alpha\lambda_\alpha S_q(\rho_\alpha) \leq S_q(\rho) \leq
\sum_\alpha\lambda_\alpha S_q(\rho_\alpha) + S(\lambda).
\end{equation}
\end{proposition}

\begin{definition}
Let $\mu$ and $\nu$ be two probability measures on
the set $I$, $\#(I)<\infty$. The relative entropy of $\mu$
with respect to $\nu$ is
\[
 S(\mu\,|\,\nu) := \sum_{i\in I}\mu_i\log\frac{\mu_i}{\nu_i}.
\]
Let $\rho_1$ and $\rho_2$ density matrices on a
finite-dimensional Hilbert space. The quantum relative
entropy of $\rho_1$ with respect to $\rho_2$ is
\[
 S_q(\rho_1\,|\,\rho_2) := \Tr\rho_1(\log\rho_1-\log\rho_2).
\]
\end{definition}

These relative entropies decrease monotonically
under positive transformations.

\begin{proposition}[Monotonicity of relative entropy]
\label{prop:monotone}
Let $M:\ctu(J)\to\ctu(I)$ be a linear positive unital map
between the continuous functions on two finite sets
$I$ and $J$. This means, for $f\in\ctu(J)$, $i\in I$
\[
M(f)(i)=\sum_{j\in J} M_{ij}f(j)
\quad \mathrm{with} \quad
M_{ij} \geq 0, \quad
\sum_{j\in J} M_{ij} = 1.
\]
The dual of $M$ maps probability measures on $I$
to probability measures on $J$. For two probability
measures $\mu$ and $\nu$ on $I$, the relative entropy
$S$ satisfies,
\[
S(M^*\mu\,|\,M^*\nu) \leq S(\mu\,|\,\nu).
\]
For the quantum case, let $M:\algg\to\alg$ be a linear
completely positive unital map between two
finite-dimensional matrix algebras. The dual of $M$
maps states on $\alg$ to states on $\algg$. For two
density matrices $\rho_1$ and $\rho_2$ on $\alg$,
the relative entropy $S_q$ satisfies,
\[
S_q(M^*\rho_1\,|\,M^*\rho_2) \leq S_q(\rho_1\,|\,\rho_2).
\]
\end{proposition}

As a consequence of this proposition,
we have the following lemma.

\begin{lemma} \label{lemm:holevo}
Let $\rho$, $\rho_\alpha$, $\alpha\in A$ and $\#(A)<\infty$,
be density matrices on a finite-dimensional Hilbert space
and let $\lambda=\{\lambda_\alpha\,|\,\alpha\in A\}$ be
a probability measure such that
\[
\rho=\sum_{\alpha\in A}\lambda_\alpha\rho_\alpha,
\]
then,
\[
S(\mathrm{diag}(\rho))-
  \sum_\alpha\lambda_\alpha S(\mathrm{diag}(\rho_\alpha))
\leq 
  S_q(\rho)-\sum_\alpha\lambda_\alpha S_q(\rho_\alpha).
\]
\end{lemma}

\textbf{Proof} \ 
Suppose $\rho,\rho_\alpha\in\matr{k}$ and $\alpha\in A$.
Consider $M : \matr{k}\to\matr{k}\otimes\ctu(A)$ defined by
$M(A)=A\otimes\idty$. This is a completely positive unital map.
Then $M^*$ is the partial trace of $\matr{k}\otimes\ctu(A)$
to $\matr{k}$. Define,
\[
\rho_1 = \sum_\alpha\lambda_\alpha\rho_\alpha
         \otimes \delta_\alpha
\quad\mathrm{and}\quad
\rho_2 = \sum_\alpha\lambda_\alpha\mathrm{diag}(\rho_\alpha)
         \otimes \delta_\alpha.
\]
Applying the monotonicity of the quantum relative entropy,
Prop.~\ref{prop:monotone}, we obtain the inequality
we are looking for. \qed

\subsection{Deterministic dynamical systems}

The Kolmogorov-Sinai construction for the dynamical entropy
of deterministic systems, see, e.g., \cite{W}, can be cast
into this algebraic framework. With the notation of
Example~\ref{exam:determ}, let $\nu$ be a probability measure
on $X$ and $\unp=\{f_k\,|\,k\in K\}$ a partition of unity in
$L^\infty(X,\mu)$. Define a probability measure on $K$ by
$\nu\circ\unp=\{\nu(f_k)\,|\,k\in K\}$. The metric or Kolmogorov-Sinai
entropy $\h{KS}[\mu,\Theta]$ for the deterministic dynamics
$\Theta$ with respect to the invariant measure $\mu$
can be written as,
\begin{equation}
\h{KS}[\mu,\Theta] := \sup_{\shp}
  \lim_{N\to\infty}\frac{1}{N}S(\mu\circ\chi_\shp^{(N)}[\Theta]),
\label{eq:KS}
\end{equation}
where the supremum is over all partitions $\shp$ of 
the phase space $X$, Example~\ref{exam:sharp}.

\section{CNT dynamical entropy}

In the following two sections, we will generalize the
dynamical entropy (\ref{eq:KS}) for deterministic systems
to stochastic systems. Our approach will be as follows.
Different quantum dynamical entropies have been proposed
in the literature and they all have to handle, albeit implicitly,
the uncertainty generated by a quantum measurement.
As explained before, a similar problem arises in
the construction of a dynamical entropy for classical
stochastic systems. We want now to reuse the different
quantum solutions to treat the measurement uncertainty
in this classical stochastic context.

We start by examining the quantum dynamical entropy proposed
by Connes, Narnhofer and Thirring in \cite{CNT}.
In our language the basic notion is the entropy
of a partition of unity with respect to a decomposition of
the invariant measure. In particular, let
$\mu=\sum_\alpha\lambda_\alpha\mu_\alpha$ be a decomposition
of the invariant measure $\mu$ as a convex combination of
probability measures $\mu_\alpha$, $\alpha\in A$ and
$\#(A)<\infty$, with coefficients $\lambda_\alpha$,
$\lambda_\alpha\geq 0$ and $\sum_\alpha\lambda_\alpha=1$.
Let $\unp=\{f_k\}$ be a partition of unity. We define
\begin{equation} \label{eq:Hrel}
 \Hrel[\mu,\{\lambda_\alpha\mu_\alpha\},\unp]
 := \sum_\alpha\lambda_\alpha
     S(\mu_\alpha\circ\unp\,|\,\mu\circ\unp)
 = S(\mu\circ\unp)
     -\sum_\alpha\lambda_\alpha S(\mu_\alpha\circ\unp).
\end{equation}
To interpret this quantity, note first that a decomposition of $\mu$
corresponds to a partition of unity $\unq=\{g_\alpha\}$ where
$\lambda_\alpha=\mu(g_\alpha)$ and $\mu_\alpha(f)=\mu(g_\alpha
f)/\mu(g_\alpha)$ for $f\in L^\infty(X,\mu)$.
In other words, the function $g_\alpha\in L^1(X,\mu)$ is
the Radon-Nikodym derivative of $\mu_\alpha$ with
respect to $\mu$.
Now $\unq$ can be seen as a finite model of $X$, whereas
$\unp$ corresponds as usual to a measurement.
Define the joint probability distribution
$\mu_{\alpha k}^{12} := \mu(g_\alpha f_k)$ with marginals
$\mu_\alpha^1=\mu(g_\alpha)$, $\mu_k^2=\mu(f_k)$.
With these definitions,
\begin{equation} \label{eq:Hrelfin}
\Hrel[\mu,\{\lambda_\alpha\mu_\alpha\},\unp]
 =S(\mu^1)+S(\mu^2)-S(\mu^{12})
\end{equation}
is the mutual information of the two marginals, or
the average amount of information obtained
about an initial condition $g_\alpha\in\unq$
by performing a measurement $f_k\in\unp$.

\subsection{CNT construction}

The entropic quantity (\ref{eq:Hrel}) is now used in
the definition of the CNT dynamical entropy of a stochastic
dynamical system. Multi-index decompositions of the measure
$\mu$ will be needed, which we write as
\[
\mu=\sum_{\oa}\lambda_{\oa}\mu_{\oa},
\]
where $\oa=(\alpha_1,\alpha_2,\ldots,\alpha_N)$,
$\alpha_n\in A_n$ and $\#(A_n)<\infty$,
$\mu_{\oa}$ are probability measures on $X$ and
$\lambda_{\oa}$ are the weights.
For every $n=1,2,\ldots,N$, the marginal of this decomposition
over all but the $n$-th index will be written as
$\mu=\sum_{\alpha_n}\lambda_{\alpha_n}^{(n)}\mu_{\alpha_n}^{(n)}$.
More explicitly,
\[
\lambda_{\beta}^{(n)} =
\sum_{\oa\,:\,\alpha_n=\beta} \lambda_{\oa}
\quad\mathrm{and}\quad
\mu_{\beta}^{(n)} = \frac{1}{\lambda_{\beta}^{(n)}}
\sum_{\oa\,:\,\alpha_n=\beta} \lambda_{\oa}\mu_{\oa}.
\]
The probability measures
$\{\lambda_{\oa}\,|\,\alpha_n\in A_n, \forall n\}$
and $\{\lambda_{\alpha_n}^{(n)}\,|\,\alpha_n\in A_n\}$
will be denoted by $\lambda$ and $\lambda^{(n)}$ respectively.
\begin{definition} \label{def:CNT}
Let $(X,\mu,\Theta)$ be a stochastic dynamical system. Define,
\begin{eqnarray}
\lefteqn{ \HH{CNT}[\mu,\{\unp_1,\unp_2,\ldots,\unp_N\}] } \nonumber \\
 &=& \sup_{\mu=\sum_{\oa}\lambda_{\oa}\mu_{\oa}}
   \Bigg( \sum_{n=1}^N
   \Hrel[\mu,\{\lambda_{\alpha_n}^{(n)}\mu_{\alpha_n}^{(n)}\},\unp_n]
   -\left(\sum_{n=1}^N S(\lambda^{(n)})-S(\lambda)\right)
   \Bigg),  \label{eq:preCNT}
\end{eqnarray}
where the supremum is over all finite $N$-index decompositions.
The CNT dynamical entropy
of $(X,\mu,\Theta)$ is given by
\begin{equation} \label{eq:CNT}
\h{CNT}[\mu,\Theta] = 
\sup_\unp \lim_{N\to\infty} \frac{1}{N}
\HH{CNT}[\mu,\{\unp,\Theta(\unp),\ldots,\Theta^{N-1}(\unp)\}].
\end{equation}
\end{definition}

Especially in the quantum case, the optimization problem in
(\ref{eq:preCNT}) is the basic obstacle for calculating
this dynamical entropy, see \cite{B, BNU, U}.
We will analyze the supremum for $N=1$ and $N=2$.

\textbf{One-time decompositions} \ 
For the case $N=1$, Eq.~(\ref{eq:preCNT}) becomes,
\begin{eqnarray*}
\HH{CNT}[\mu,\unp]
 &=& \sup_{\mu=\sum_\alpha\lambda_\alpha\mu_\alpha}
   \Hrel[\mu,\{\lambda_\alpha\mu_\alpha\},\unp] \\
 &=& S(\mu\circ\unp)
   -\inf_{\mu=\sum_\alpha\lambda_\alpha\mu_\alpha}
   \sum_\alpha\lambda_\alpha S(\mu_\alpha\circ\unp).
\end{eqnarray*}
The infimum in the second term of the right hand side
is a convex optimization problem~: find the infimum
of the concave entropy functional over the convex domain
of finite decompositions of the measure $\mu$.
This infimum will then be reached on the set of
extremal points of this domain.

Assume that the partition of unity $\unp=\{f_k\,|\,k\in K\}$
consists of simple functions. The result for general $f_k$
will follow by continuity. For such simple functions
their exists a partition $\shp=\{C_i\,|\,i\in I\}$ of X,
$\#(I)\leq\infty$, such that $f_k=\sum_i f_{ik}\chi_{C_i}$.
The measures $\mu$, $\mu_\alpha$ can then be considered
as measures on the finite set $I$,
$\mu=\{\mu(C_i)\,|\,i\in I\}$ and
$\mu_\alpha=\{\mu_\alpha(C_i)\,|\,i\in I\}$.

Let us now determine the extremal finite decompositions of
the measure $\mu$, or, equivalently, the extremal probability
measures on $I\times A$ with $\#(A)<\infty$ such that
the marginal over the $\alpha$-index equals $\mu$.
We claim that these extremal measures are $\mu_f$,
characterized by a map $f:I\to A$ such that
$\mu_f(i,\alpha)=\delta_{\alpha,f(i)}\mu(C_i)$. Indeed,
$\mu_f$ is a probability measure on $I\times A$,
\[
\sum_{i,\alpha}\mu_f(i,\alpha)
 = \sum_{i,\alpha}\delta_{\alpha,f(i)}\mu(C_i)
 = \sum_{i}\mu(C_i) = 1.
\]
The measure $\mu_f$ has $\mu$ as marginal over the
$\alpha$-index,
\[
\sum_{\alpha\in A}\mu_f(i,\alpha)
 = \sum_{\alpha\in A}\delta_{\alpha,f(i)}\mu(C_i)
 = \mu(C_i).
\]
Moreover, $\mu_f$ is extremal. Suppose we can write $\mu_f$
as a convex combination of $\nu_1$ and $\nu_2$,
two probability measures on $I\times A$ with marginal $\mu$,
\[
\mu_f=\frac{1}{2}\nu_1+\frac{1}{2}\nu_2.
\]
Substituting the explicit form for $\mu_f$, one immediately
gets $\mu_f=\nu_1=\nu_2$. Finally, all the extremal points are
of this form because every probability measure on $I\times A$
with marginal $\mu$, can be written as a convex combination,
\[
\mu_\alpha(C_i)=\sum_f c_f\mu_f(i,\alpha),
\]
with $c_f\geq 0$ and $\sum_fc_f=1$.

The infimum will thus be reached on this set of measures
$\mu_f$. Moreover, we can restrict our attention to injective
maps $f$. This follows again by the concavity of the entropy
functional. The order of the indices $\alpha$ is of no
importance, so we can take as optimal decomposition
$\mu_\alpha=\delta_i$ and $\lambda_\alpha=\mu(C_i)$.
As a result, for simple functions $f_k$,
\[
\HH{CNT}[\mu,\unp] = S(\mu\circ\unp) -
 \sum_i \mu(C_i) S(\{f_{ik}\,|\,k\in K\}),
\]
or, for general $f_k$,
\begin{equation} \label{eq:Hrelsol}
\HH{CNT}[\mu,\unp]
 = S(\mu\circ\unp) - \int d\mu(x)S(\delta_x\circ\unp)
 = \int d\mu(x)S(\delta_x\circ\unp\,|\,\mu\circ\unp).
\end{equation}

\textbf{Two-times decompositions} \ 
In contrast with the case $N=1$, the optimization problem
(\ref{eq:preCNT}) is {\em not} convex for the case $N=2$.
To see this, suppose the partitions of unity
$\unp=\{f_k\,|\,k\in K\}$ and $\unq=\{g_l\,|\,l\in L\}$
consist of simple functions,
$f_k=\sum_i f_{ik}\chi_{C_i}$ and $g_l=\sum_j g_{jl}\chi_{D_j}$.
Here, $\shp=\{C_i\}$ and $\shq=\{D_j\}$ are finite partitions
of $X$ and so is $\shp\lor\shq=\{C_i\cap D_j\}$.

Define now a probability measure on the composed sytem
$A\times B\times I\times J\times K\times L$,
\[
\mu_{\alpha\,\beta\,i\,j\,k\,l}^{1\,2\,3\,4\,5\,6} =
\lambda_{(\alpha,\beta)}\mu_{\alpha\beta}(C_i\cap D_j)f_{ik}f_{jl}.
\]
This probability measure has marginals
\[
\mu_k^5=\mu(f_k), \quad \mu_l^6=\mu(g_l), \quad
\mu_{\alpha\beta}^{12}=\lambda_{(\alpha,\beta)},
\]
\[
\mu_{\alpha k}^{15}=\lambda_\alpha^{(1)}\mu_\alpha^{(1)}(f_k), \quad
\mu_{\beta l}^{26}=\lambda_\beta^{(2)}\mu_\beta^{(2)}(g_l).
\]
The functional to optimize can then be written as
\[
S(\mu^5)+S(\mu^6)+S(\mu^{12})-S(\mu^{15})-S(\mu^{26}).
\]
The task is now to optimally couple subsystems $A$ and $B$
with $I$ and $J$. If this optimization problem were convex,
the optimal coupling would identify $A$ with $I$ and
$B$ with $J$, as in the case $N=1$. This decomposition can
yield a negative value for the supremum. However,
the supremum has to be positive because the functional is zero
for the trivial decomposition, i.e., $\#(A)=1$ and $\#(B)=1$.
We conclude that this optimization problem is not convex.

\subsection{Hudetz construction}

The CNT construction seems to be intractable because of
the supremum over all multi-index decompositions of the
invariant measure, Eq.~(\ref{eq:preCNT}).
These multi-index decompositions were introduced to obtain
finite-dimensional algebras for the one-time restrictions.
For stochastic systems this algebraic structure is absent anyway.
The following construction appears to be more natural.

\begin{definition} \label{def:Hud}
Let $(X,\mu,\Theta)$ be a stochastic dynamical system. Define,
\begin{equation} \label{eq:preHud}
 \HH{Hud}[\mu,\unp] := \sup_{\mu=\sum_\alpha\lambda_\alpha\mu_\alpha}
 \Hrel[\mu,\{\lambda_\alpha\mu_\alpha\},\unp].
\end{equation}
The Hudetz (Hud) dynamical entropy of $(X,\mu,\Theta)$
is given by
\begin{equation} \label{eq:Hud}
 \h{Hud}[\mu,\Theta] := \sup_\unp \limsup_{N\to\infty}
  \frac{1}{N} \HH{Hud}[\mu,\unp^{(N)}[\Theta]].
\end{equation}
\end{definition}

The optimization problem at any time $N$ is now the same
as the one encountered in the CNT construction for one time
decompositon. This supremum was worked out explicitly,
Eq.~\ref{eq:Hrelsol},
\begin{eqnarray} \label{eq:Hhudsol}
\HH{Hud}[\mu,\unp]
 = S(\mu\circ\unp)-\int\!d\mu(x)\,S(\delta_x\circ\unp)
 = \int\!d\mu(x)S(\delta_x\circ\unp\,|\,\mu\circ\unp)
\end{eqnarray}
Moreover, restricting the supremum to partitions of unity $\unp$
which correspond to sharp measurements, leads to the same result.

\begin{proposition} \label{prop:Hudsharp}
For a stochastic dynamical system $(X,\mu,\Theta)$ holds
\[
\h{Hud}[\mu,\Theta] = \sup_\shp \limsup_{N\to\infty}
  \frac{1}{N} \HH{Hud}[\mu,\chi_\shp^{(N)}[\Theta]].
\]
\end{proposition}

\textbf{Proof} \ 
We have to show that for every partition of unity $\unp=\{f_k\}$
there exist a partition $\shp=\{C_i\}$ of $X$ such that,
\[ 
\HH{Hud}[\mu,\unp^{(N)}[\Theta]] \leq
\HH{Hud}[\mu,\chi_\shp^{(N)}[\Theta]].
\]
We consider the case $N=2$. The proof for other $N$ is analogous.

Assume that $f_k$ are simple functions. The result for general $f_k$ will
follow by continuity. For simple functions $f_k$ there exists a partition
$\shp=\{C_i\}_i$ of $X$ such that $f_k=\sum_i f_{ik}\chi_{C_i}$ with
$M=[f_{ik}]$ a stochastic matrix. Element $(k_0,k_1)$ of the refined partition
$\unp^{(2)}[\Theta]=\unp\lor\Theta(\unp)$ is then
\[
f_{k_0}\Theta(f_{k_1})  = \sum_{i_0i_1} f_{i_0k_0}f_{i_1k_1}
  \chi_{C_{i_0}}\Theta(\chi_{C_{i_1}})
\]
Thus,
\[
\mu\circ\unp^{(2)}[\Theta]
 = (M\otimes M)^*(\mu\circ\chi_\shp^{(2)}[\Theta])
\]
with $M\otimes M$ a positive unital transformation.
By Prop.~\ref{prop:monotone} one obtains,
\[
\HH{Hud}[\mu,\unp^{(2)}[\Theta]]\leq
\HH{Hud}[\mu,\chi_\shp^{(2)}[\Theta]].
\] \qed

Referring to the information-theoretic interpretation of
$\Hrel[\mu,\{\lambda_\alpha\mu_\alpha\},\unp]$, the quantity
$\HH{Hud}[\mu,\unp]$ can be seen as the mutual information
between the initial state and the measurement outcomes
for the best model of the state space $X$, namely $X$ itself.
Therefore, $\HH{Hud}[\mu,\unp]$ equals the information obtained
about an initial state $x\in X$ by performing a measurement
$f_k\in\unp$. The Hudetz dynamical entropy $\h{Hud}[\mu,\Theta]$
equals the average information gain. In this way, out of the three
sources of dynamical entropy, only the system dynamics contributes.

\begin{example}[Deterministic systems]
By Prop.~\ref{prop:Hudsharp}, we can restrict our
attention to sharp measurements. Recall that deterministic
dynamics transform sharp measurements into sharp ones.
Therefore, all the probability measures $\delta_x\circ\unp$
appearing in (\ref{eq:Hhudsol}) are pure and $\HH{Hud}[\mu,\unp]=S(\mu\circ\unp)$. We conclude that
$\h{Hud}=\h{KS}$ for deterministic systems.
\end{example}

\begin{example}[Finite systems]
Consider a dynamical system with a finite state space $X$.
Denote the invariant measure by $\mu=\{\mu_x\,|\,x\in X\}$.
By Eq.~\ref{eq:Hhudsol},
\[
\HH{Hud}[\mu,\unp]
 = \sum_{kx} \mu_x f_k(x) \log \frac{f_k(x)}{\mu(f_k)}
 \leq \sum_{kx} \mu_x f_k(x) \log \frac{1}{\mu_x}
 = S(\mu).
\]
This quantity is finite and does not depend on $N$ when
$\unp$ is replaced by $\unp^{(N)}[\Theta]$ in (\ref{eq:Hud}).
We conclude that $\h{Hud}=0$ for finite systems.
\end{example}

Note that the CNT dynamical entropy gives the same
results for these two examples. Finally, let us compare
the Hudetz dynamical entropy with another definition for
the dynamical entropy of stochastic systems.
It is closely related to \cite{M}, but as explained before,
we use another refinement of partitions,
(\ref{eq:refineAF}) instead of (\ref{eq:refineMak}).
First, a density matrix is constructed,
\begin{equation} \label{eq:rhoMak}
 \left(\ro{Mak}[\mu,\unp]\right)_{kl} := \mu(\sqrt{f_kf_l}).
\end{equation}
The Makarov dynamical entropy is given by
\begin{equation} \label{eq:Mak}
 \h{Mak}[\mu,\Theta] := \sup_\unp \limsup_{N\to\infty}
 \frac{1}{N} S_q(\ro{Mak}[\mu,\unp^{(N)}[\Theta]]).
\end{equation}

This dynamical entropy leads to the same result as the Hudetz
dynamical entropy for the two examples discussed. Moreover,
it holds that $\h{Hud}\leq\h{Mak}$. Indeed, from Lemma~\ref{lemm:holevo},
\[
S(\mathrm{diag}(\rho))-\sum_i\lambda_iS(\mathrm{diag}(\rho_i))
\leq S_q(\rho).
\]
This is equivalent with $\HH{Hud}[\mu,\unp]\leq S_q(\ro{Mak}[\mu,\unp])$
for all partitions of unity $\unp$.

\section{AFL dynamical entropy}

In \cite{AF1} another quantum dynamical entropy was proposed, based on an idea
of Lindblad,  by mapping the evolution of a dynamical system onto a quantum
spin chain. For the classical stochastical systems we are interested in, the
definition is as follows.

\begin{definition} \label{def:AF}
Let $(X,\mu,\Theta)$ be a stochastic dynamical system. Define
the density matrix $\ro{AFL}\,^{(N)}$ by
\begin{equation} \label{eq:rhoAF}
 \left(\ro{AFL}\,^{(N)}[\mu,\Theta,\unp]\right)_{\ok,\ol} :=
 \mu \Big( \sqrt{f_{k_0}f_{l_0}}\Theta\Big(\sqrt{f_{k_1}f_{l_1}}
\ldots \Theta\Big(\sqrt{f_{k_{N-1}}f_{l_{N-1}}}\Big)\Big) \Big).
\end{equation}
The AFL dynamical entropy of $(X,\mu,\Theta)$
is given by
\begin{equation} \label{eq:AF}
 \h{AFL}[\mu,\Theta] := \sup_\unp \limsup_{N\to\infty} \frac{1}{N}
S_q(\ro{AFL}\,^{(N)}[\mu,\Theta,\unp]).
\label{eq:AF1}
\end{equation}
\end{definition}

Note that the density matrix $\ro{AFL}\,^{(N)}[\mu,\Theta,\unp]$
is different from $\ro{Mak}[\mu,\unp^{(N)}[\Theta]]$ for $N>2$.
For sharp measurements $\chi_\shp$ with $\shp$ a partition of $X$,
the density matrix $\ro{AFL}\,^{(N)}[\mu,\Theta,\chi_\shp]$ is
diagonal. In that case,
\[
S_q(\ro{AFL}\,^{(N)}[\mu,\Theta,\unp])
 = S(\mu\circ\unp^{(N)}[\Theta]).
\]
As a consequence,
\[
\h{AFL}[\mu,\Theta] \geq \sup_{\unp\ \mathrm{sharp}}
\limsup_{N\to\infty} \frac{1}{N} S(\mu\circ\unp^{(N)}[\Theta])
\]
In fact, equality holds.

\begin{proposition} \label{prop:AFsharp}
For a stochastic dynamical system $(X,\mu,\Theta)$ holds
\[
\h{AFL}[\mu,\Theta] = \sup_\shp \limsup_{N\to\infty}
  \frac{1}{N} S(\mu\circ\chi_\shp^{(N)}[\Theta]).
\]
\end{proposition}

\textbf{Proof} \ 
We have to show that for every partition of unity $\unp=\{f_k\}$
there exist a partition $\shp=\{C_i\}$ of $X$ such that,
\[
S(\ro{AFL}\,^{(N)}[\mu,\Theta,\unp])
 \leq S(\mu\circ\chi_\shp^{(N)}[\Theta]).
\]
We consider the case $N=2$. The proof for other $N$ is analogous.

Assume that $f_k$ are simple functions.
The result for general $f_k$ will follow by continuity.
For such simple functions, there exists a partition of $X$,
$\shp=\{C_i\,|\,i\in I\}$ and $\#(I)<\infty$, such that
$f_k=\sum_i f_{ik}\chi_{C_i}$ with $[f_{ik}]$ a stochastic
matrix. Component $(k_0,k_1),(l_0,l_1)$ of the density
matrix $\ro{AFL}\,^{(2)}[\mu,\Theta,\unp]$ is
\begin{eqnarray*}
\lefteqn{ \mu \left( \sqrt{f_{k_0}f_{l_0}}
  \Theta\Big(\sqrt{f_{k_1}f_{l_1}}\Big) \right) } \\
&=& \mu \left( \sqrt{\sum_{i_0j_0}f_{i_0k_0}f_{j_0l_0}
  \chi_{C_{i_0}}\chi_{C_{j_0}}}
  \Theta\Bigg(\sqrt{\sum_{i_1j_1}f_{i_1k_1}f_{j_1l_1}
  \chi_{C_{i_1}}\chi_{C_{j_1}}}\Bigg) \right) \\
&=& \mu \left( \sqrt{\sum_{i_0}f_{i_0k_0}f_{i_0l_0}
  \chi_{C_{i_0}}}
  \Theta\Bigg(\sqrt{\sum_{i_1}f_{i_1k_1}f_{i_1l_1}
  \chi_{C_{i_1}}}\Bigg) \right) \\
&=& \sum_{i_0i_1} \mu \left( \chi_{C_{i_0}}
  \Theta\left(\chi_{C_{i_1}}\right) \right)
  \sqrt{f_{i_0k_0}f_{i_0l_0}} \sqrt{f_{i_1k_1}f_{i_1l_1}}
\end{eqnarray*}
This a convex combination of $\#(I)^2$ vector states.
The coefficient of term $(i_0,i_1)$ is
$\mu(\chi_{C_{i_0}}\Theta(\chi_{C_{i_1}}))$
and component $(k_0,k_1)$ of the corresponding vector is $\sqrt{f_{i_0k_0}f_{i_1k_1}}$.
These vectors are normalized because $[f_{ik}]$ is stochastic.
Applying the second inequality in (\ref{eq:concave})
finishes the proof. \qed

Prop.~\ref{prop:AFsharp} leads to the following interpretation
of $\h{AFL}[\mu,\Theta]$. It is the average uncertainty on the outcome
of sharp measurements. Out of the three sources of dynamical entropy,
both the system dynamics and the stochasticity contribute.

\begin{example}[Deterministic systems]
By comparing Eq.~\ref{eq:KS} and Prop.~\ref{prop:AFsharp},
$\h{AFL} = \h{KS}$ for deteministic systems.
\end{example}

\begin{example}[Finite systems]
The dynamical entropy $\h{AFL}$ can be strictly positive
for finite systems. In this case, the supremum over
all sharp partitions is reached for the extremal partition,
i.e., $\shp=\{\{x\}\,|\,x\in X\}$.
For a Bernoulli process with probabilities $\{p_x\}$ and
$\sum_xp_x=1$, one obtains $\h{AFL}=\sum_x\eta(p_x)$,
whereas $\h{Hud}=0$.
\end{example}

Finally, we compare the AFL dynamical entropy with
the other definitions. The given interpretation and the finite
case example suggest the inequality $\h{Hud}\leq\h{AFL}$.
This can be easily proven by Lemma~\ref{lemm:holevo}.
Another definition for the dynamical entropy of stochastic
systems was given in \cite{KOW}, based on \cite{AOW} for
deterministic systems. In our notation it can be written as,
\[
\h{KOW}[\mu,\Theta] = \sup_{\unp} \limsup_{N\to\infty}
\frac{1}{N} S(\mu\circ\unp^{(N)}[\Theta]).
\]
By comparing this with Prop.~\ref{prop:AFsharp}, $\h{AFL}\leq\h{KOW}$.
Because unsharp measurements are allowed, they can contribute to the dynamical
entropy. The three sources of dynamical entropy are thus taken into account. As
a consequence, $\h{KS}\leq\h{KOW}$ for deterministic systems where strict
inequality can hold. Even stronger, without restricting the set of allowed
measurements, this dynamical entropy will always be infinite. Indeed, the
partition of unity consisting of $k$ elements $\frac{1}{k}\idty$ leads to a
dynamical entropy $\log k$. This can grow without bounds.

\textbf{Acknowledgements} \ 
It is a pleasure to acknowledge constructive discussions with R.~Alicki who
pointed out reference~\cite{M} to us.

\end{document}